\documentclass[]{elsarticle}

\usepackage{lineno,hyperref}
\modulolinenumbers[5]
\usepackage{subfigure}
\journal{Journal}
\usepackage{amsmath,amssymb}
\usepackage{lmodern}
\usepackage[margin=1in]{geometry}
\usepackage{subfigure}

\usepackage{graphicx}

\begin{document}

\begin{frontmatter}

\title{Electrical frequency discrimination by fungi \textit{Pleurotus ostreatus}}

\author[1]{Dawid Przyczyna}
\author[1]{Konrad Szacilowski}
\author[2,3]{Alessandro Chiolerio}
\author[3,4]{Andrew Adamatzky}
\address[1]{Academic Centre for Materials and Nanotechnology, AGH University of Science and Technology, Krakow, Poland}
\address[2]{Istituto Italiano di Tecnologia, Center for Converging Technologies, Soft Bioinspired Robotics, Via Morego 30, 16165 Genova, Italy}
\address[3]{Unconventional Computing Lab, UWE, Bristol, UK}
\address[4]{Department of Electrical and Computer Engineering, Democritus University of Thrace, Xanthi, Greece}

\begin{abstract}
We stimulate mycelian networks of oyster fungi \textit{Pleurotus ostreatus} with low frequency sinusoidal electrical signals. We demonstrate that the fungal networks can discriminate between frequencies in a fuzzy or threshold based manner. Details about the mixing of frequencies by the mycelium networks are provided. The results advance the novel field of fungal electronics and pave ground for the design of living, fully recyclable, electron devices.
\end{abstract}


\end{frontmatter}


\begin{abstract}
We stimulate mycelian networks of oyster fungi \textit{Pleurotus ostreatus} with low frequency sinusoidal electrical signals. We demonstrate that the fungal networks can discriminate between frequencies in a fuzzy-like or threshold based manner. Details about the mixing of frequencies by the mycelium networks are provided. The results advance the novel field of fungal electronics and pave ground for the design of living, fully recyclable, electron devices.
\end{abstract}

\section{Introduction}

Fungal electronics aims to design bio-electronic devices with living networks of fungal mycelium~\cite{adamatzky2021fungal} and proposes novel and original designs of information and signal processing systems. The reasons for developing fungal electronic devices are following. Mycelium bound composites (grain or hemp substrates colonised by fungi) are environmentally sustainable growing bio-materials~\cite{karana2018material,jones2020engineered,cerimi2019fungi}. They have been already used in insulation panels~\cite{pelletier2013evaluation,elsacker2020comprehensive,dias2021investigation,wang2016experimental,cardenas2020thermal}, packaging materials~\cite{holt2012fungal,mojumdar2021mushroom}, building materials and architectures~\cite{adamatzky2019fungal} and wearables~\cite{adamatzky2021reactive,silverman2020development,karana2018material,appels2020use,jones2020leather}. To make the fungal materials functional we need to embed flexible electronic devices into the materials. Hyphae of fungal mycelium spanning the mycelium bound composites can play a role of unconventional electronic devices. interestingly, their topology is very similar to conducting polymer dendrites \cite{leo, filament1}. These properties originate not only from common topology \cite{pismen1} but also from complex electron transport phenomena.Therefore, it is not surprising that electrical properties of mycelial hyphae and conducting polymer filaments have similar electrical properties: proton hopping and ionic transport in hyphae \textit{vs} ionic and electronic transport in polymers. Such transport duality must result in highly nonlinear voltage/current characteristics, which in turn, upon AC stimulation must result in generation of complex Fourier patterns in resulting current, as well as other phenomena relevant from the point of view of unconventional computing, e.g. stochastic resonance \cite{kasai}.  

We have already demonstrated that we achieved in implementing memristors~\cite{beasley2022mem}, oscillators~\cite{adamatzky2021electrical}, photo-sensors~\cite{beasley2020fungal}, pressure sensors~\cite{adamatzky2022living}, chemical sensors~\cite{dehshibi2021stimulating} and Boolean logical circuits~\cite{roberts2021mining} with living mycelium networks. Due to nonlinear electric response of fungal tissues, they are ideally suited for transformation of low-frequency AC signals. This paper is devoted to frequency discriminators and transformers, which are a significant contribution to the field of fungal electronics.   

Electrical communication in mycelium networks is an almost unexplored topic.
Fungi exhibit oscillations of extracellular electrical potential, which can be recorded via differential electrodes inserted into a substrate colonised by mycelium or directly into sporocarps~\cite{slayman1976action,olsson1995action,adamatzky2018spiking}. In experiments with recording of electrical potential of oyster fungi \emph{Pleurotus djamor} we discovered two types of spiking activity:  high-frequency 6~mHz and low-freq 1~mHz~\cite{adamatzky2018spiking} ones. While studying other species of fungi, \emph{Ganoderma resinaceum}, we found that the most common signature of an electrical potential spike is 2-3~mHz~\cite{adamatzky2021electrical}. In both species of fungi we observed bursts of spikes within trains of impulses similar to that observed in animal central nervous system~\cite{cocatre1992identification,legendy1985bursts}.  In~\cite{dehshibi2021electrical} we demonstrated that information-theoretical complexity of fungal electrical activity exceeds the complexity of European languages. In \cite{adamatzky2022language} we analysed the electrical activity of  \emph{Omphalotus nidiformis}, \emph{Flammulina velutipes}, \emph{Schizophyllum commune} and  \emph{Cordyceps militaris}. We assumed that the spikes of electrical activity could be used by fungi to communicate and process information in mycelium networks and demonstrated that distributions of fungal word lengths match that of human languages. Taking all the above into account it would be valuable to analyse the electrical reactions of fungi to strings of electrical oscillations, featuring frequencies matching those of the supposed fungal language. 
The present paper advances our research and development in (1) fungal electronics and (2) communication in mycelium networks by proposing novel and original designs of frequency discriminators based on living fungi.

\section{Methods}

\begin{figure}[!tbp]
\centering
\includegraphics[width=0.2\textwidth]{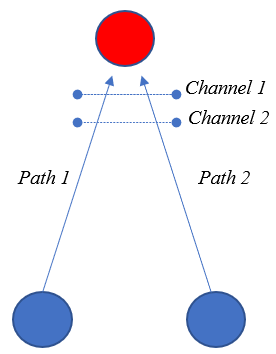}
\caption{Diagram showing connecting points to the fungi sample. The blue circles represent two places for the input signal injection whereas the red circle represents the ground.}
\label{fig01}
\end{figure}

 A slab of substrate, 200~g, colonised by \textit{Pleurotus ostreatus} (Ann Miller's Speciality Mushrooms, UK, \url{https://www.annforfungi.co.uk/shop/oyster-grain-spawn/}) was placed at the bottom of a 5~l plastic container. Measurements were performed in a classic two electrode setup. Electric contacts to the fungi sample were made using iridium-coated stainless steel sub-dermal needle electrodes (purchased by Spes Medica S.r.l., Italy), with twisted cables. Signal was applied with  4050B Series Dual Channel Function/Arbitrary Waveform Generators (B\&K Precision Corporation). Signals featuring a series of frequencies --- 1-10~mHz with a 1~mHz step and 10-100~mHz with a 10~mHz step --- have been applied between two points of the fungi and measured with two differential channels on ADC-24 (purchased by Pico Technology, UK) high-resolution data logger with a 24-bit analog-to-digital converter. We have chosen these particular intervals of frequencies because they well cover frequencies of action-potential spiking behaviour of a range of fungi species~\cite{adamatzky2018spiking,adamatzky2021electrical,adamatzky2022language}.
 
 For these frequencies, the sinusoidal signal was applied along two paths separately. Finally, mixing of signals was performed for 1~mHz base frequency applied on Path 1 and a series of frequencies on the Path 2. Frequencies used on Path 2 are 2, 5 and 7  mHz). Fast Fourier transform (FFT) was calculated with Origin Pro software. Blackman window function was used as it is best suitable for the representation of amplitudes~\cite{dactron2003understanding}.
 Fuzzy sets for inference of new input data were constructed using "fuzzylogic 1.2.0" Python package.

\section{Results}
\label{results}

\begin{figure}[!tbp]
\includegraphics[width=\textwidth]{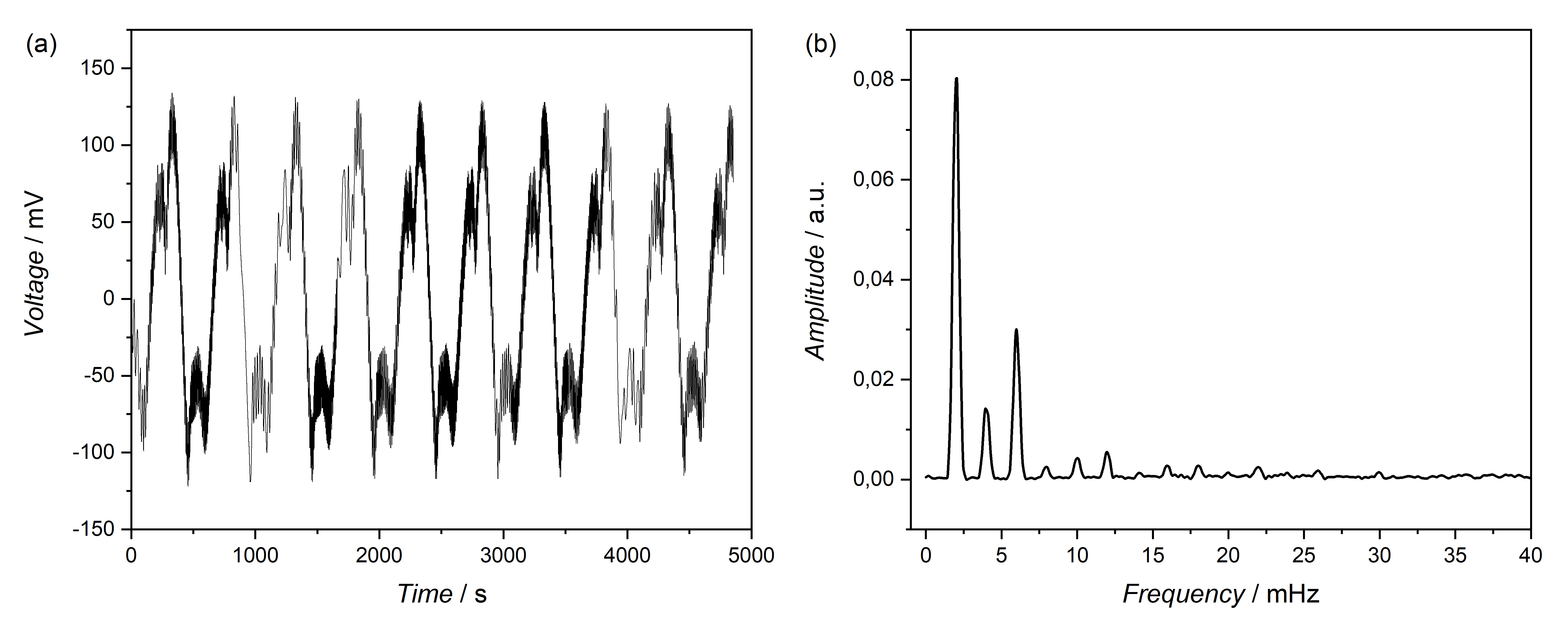}
\caption{
Exemplary response of the fungal sample to 2~mHz, 10~Vpp sinusoidal electrical stimulation (a) and FFT for the same response.}
\label{fig02}
\end{figure}

A response of the fungi sample to electrical stimulation is shown in Fig~\ref{fig01}a. In all measurements, electrical activity with frequency 50-200~mHz was observed even when substrates were not stimulated. This activity is attributed to endogenous oscillations of electrical potential of fungi~\cite{adamatzky2018spiking,adamatzky2021electrical,adamatzky2022language}.

\begin{figure}[!tbp]
\includegraphics[width=\textwidth]{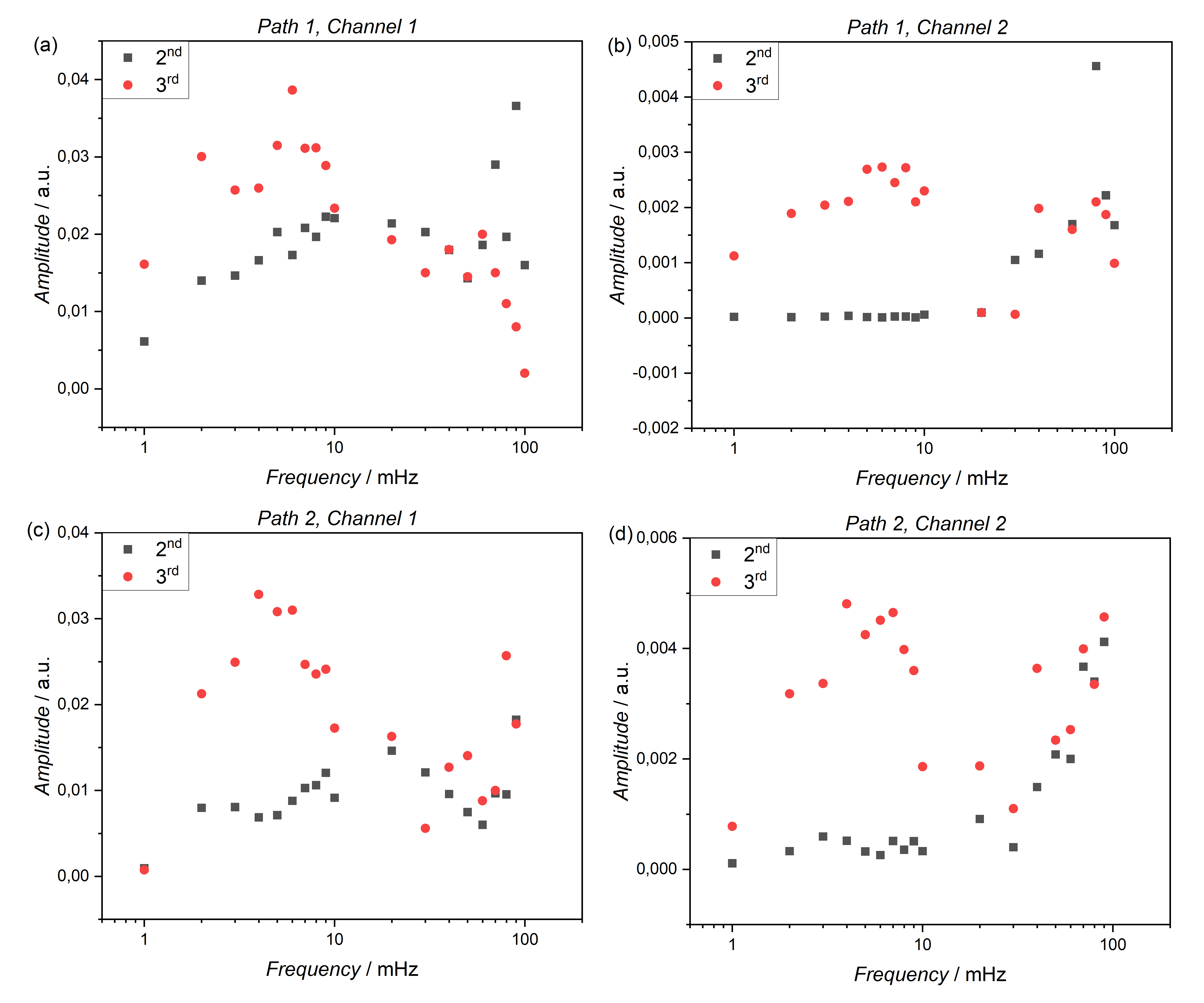}
\caption{Collection of 2\textsuperscript{nd} and
3\textsuperscript{rd} harmonic amplitudes obtained for the measured fungi response, for two signal paths and two differential channels.}
\label{fig03}
\end{figure}

Exemplary generations of higher harmonics are shown in Fig.~\ref{fig02}b. In some cases presented on Fig~\ref{fig03},
2\textsuperscript{nd} harmonic is more damped than the 3\textsuperscript{rd} harmonic. Generally, for frequencies below 10~mHz,
higher amplitudes were observed for 3\textsuperscript{rd} harmonic
versus the 2\textsuperscript{nd}.

\begin{figure}[!tbp]
\includegraphics[width=\textwidth]{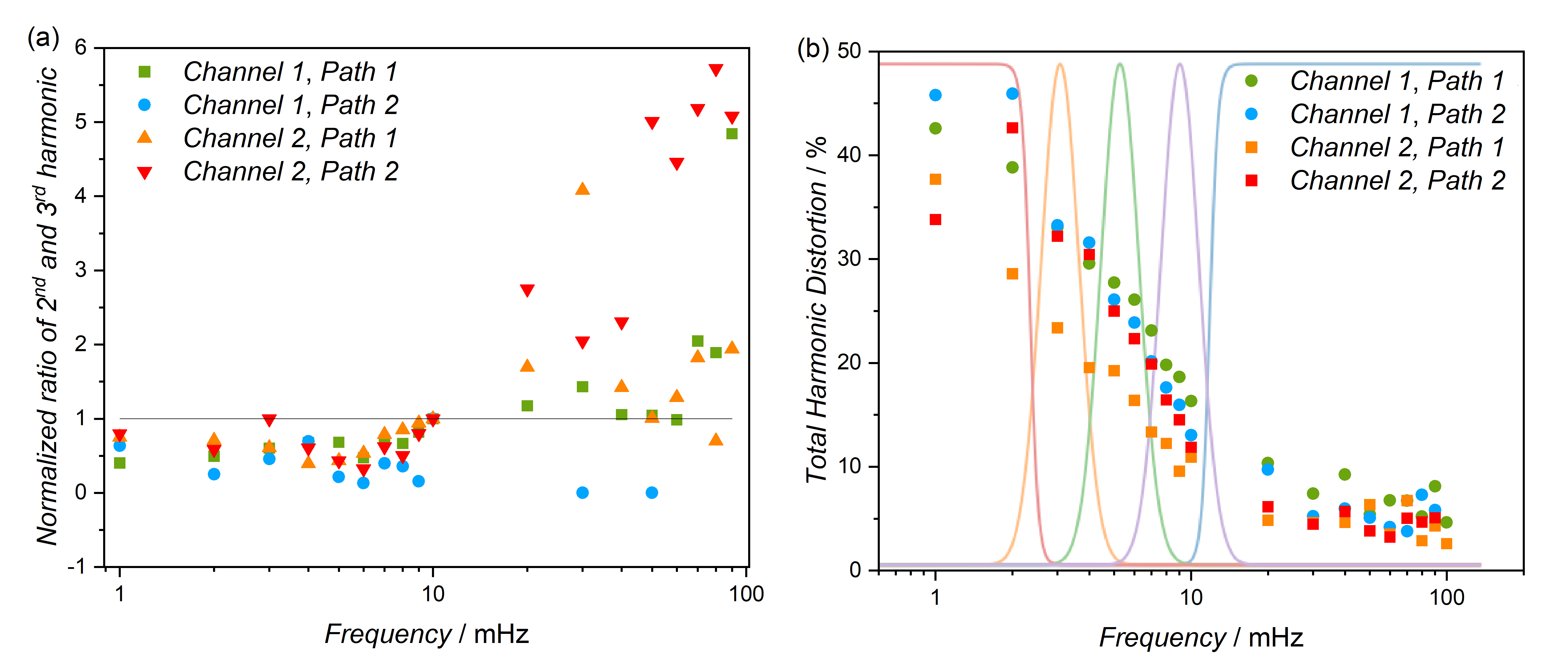}
\caption{Harmonic distributions.
(a)~Normalised ratios of 2\textsuperscript{nd} vs 3\textsuperscript{rd} harmonics for analysed signals. Straight line marks threshold frequency.
(b)~Total Harmonic Distortion calculated for fungi sample. Proposition of the fuzzy sets is included in the background.}
\label{fig04}
\end{figure}

The ratio of the 2\textsuperscript{nd} to 3\textsuperscript{rd} harmonic amplitudes was calculated to better illustrate the changes between them (Fig~\ref{fig04}a). The calculated ratios were then normalised to the ratio of harmonics at 10mHz. Points at 30 and 50~mHz in 1 path, and 2 channel were treated as outliers because the ratios at these frequencies were disproportionally larger than those at other frequencies, which disturbed data visualisation. Besides, the omitted data points in the presented graph still support the observation that in general, below 10~mHz, the ratio of the 2\textsuperscript{nd} and 3\textsuperscript{rd} harmonics are smaller than for higher frequencies. 

In the next step, Total Harmonic Distortion (THD) of the measured signal
was calculated (Fig~\ref{fig04}b). THD is the ratio between the fundamental frequency amplitude $V_0$ and the amplitude of higher harmonics $V_n$:

\begin{equation}
\text{THD}_{F} = \frac{\sqrt{V_{2}^{2} + V_{3}^{2} + V_{4}^{2} + \ldots}}{V_{1}}
\label{eq01}
\end{equation}

where \emph{V\textsubscript{n}} is the \emph{n}th amplitude of the
frequency of successive higher harmonic peaks observed in the Fourier
spectra. Furthermore, normalisation to 100\% of the THD parameter can be
applied as follows:

\begin{equation}
\text{THD}_{R} = \frac{\text{TH}D_{F}}{\sqrt{1 + \text{THD}_{F}^{2}}},
\end{equation}
where $R$ in \emph{THD\textsubscript{R}} stands for ``root mean square".

For the frequencies below 10~mHz, higher values of THD (up to 45.9\%) can be observed in relation to higher frequencies, which tend to exhibit
lower THD values (below 10\%). The THD of a pure signal ranges between different values, for example a square wave features a THD of 48.3\% and a triangular wave features a THD of 12.1\%. This result may suggest changes in the dominant conductivity type: slower signals are more distorted and faster signals are much less distorted. Lower THD values are obtained, when the generation of higher harmonics of the modulated signal is low, hence the fungi sample has lower effect on its transformation. This effect is a consequence of a dual electric charge transport mechanism in mycelium. Furthermore, the changes occurring at low frequencies indicate, that slow physical phenomena (as diffusion) are critically responsible for the distortion of electric signals. This effect is similar to those observed in the case of solid-state memristor, however in the latter case the dependence is opposite \cite{przyczyna2022}. It can be concluded that in the studied case at high frequencies only one, faster conductivity mode plays a significant role. Therefore, the nonlinear character of electric transport is much less pronounced and signal can apparently ``fly through'' the sample and can be transmitted across a macroscopic distance with low distortion.

As the changes of THD parameter below 10mHz occurs in a rather continuous manner, arbitrary linguistic (very low, low, medium, high, very high, etc.) could be defined for ranges of obtained values. Following, membership function could be specified for the allocation of data into sets so that fuzzification of data could be implemented and allow for inference of given new input data into proper category. \cite{mendel} Proposition for such sets is depicted in the background of Fig~\ref{fig04}b. Two sigmoidal sets were selected for the boundary and three Gaussian sets for the center of the data.

The results demonstrate that, based on increase of the THD parameter or on the amplitude values of 2\textsuperscript{nd} and 3\textsuperscript{rd} harmonic components, signal discrimination based on its frequency could be realised. 

\begin{figure}[!tbp]
\includegraphics[width=0.9\textwidth]{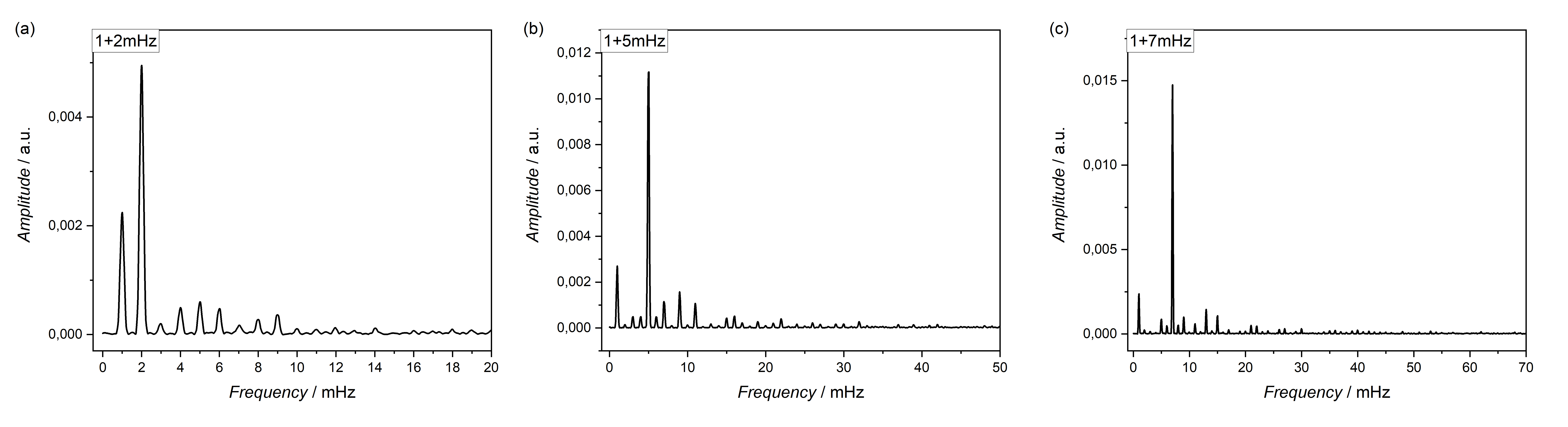}
\caption{Result of frequency mixing in the fungi samples. For each measurement, base 1~mHz driving signal was used on Path 1 (Fig.~\ref{fig01}). For each successive measurement, higher frequency signal was applied to the Path 2.}
\label{fig05}
\end{figure}

After analysis of single signal paths, signals were applied to the two signal paths at once. Results show that with increasing frequency,
further damping of the 2\textsuperscript{nd} harmonic is achieved. Furthermore, satellite frequencies appear around base frequencies as well as around higher harmonics. For example, on the Fig.~\ref{fig05}b, for the mixing of 1 and 5~mHz signal, higher frequencies --- 9~mHz and 11~mHz --- around damped 10~mHz 2\textsuperscript{nd} harmonic are present. This effect is present as well for the 1 mHz and 7 mHz  mixed frequencies. The results indicate a nontrivial frequency mixing scheme, which may results in vermicular transport phenomena within percolated, highly branched network of mycelial hyphae. 

\section{Conclusion}

We demonstrated that fungal mycelium networks modify frequencies of external electrical inputs. Damping of 2\textsuperscript{nd} harmonic and amplification of the 3\textsuperscript{nd} harmonic amplitudes below 10mHz allow for frequency discrimination in a threshold manner. The frequency discrimination could occur in a continuous manner,with the help of the concepts of fuzzy logic based on THD parameter.

\section{Acknowledgement}

DP has been partly supported by the EU Project POWR.03.02.00-00-I004/16.
AA was supported by the funding from the European Union's Horizon 2020 research and innovation programme FET OPEN ``Challenging current thinking'' under grant agreement No 858132. 



\begin{thebibliography}{99}
\expandafter\ifx\csname url\endcsname\relax
  \def\url#1{\texttt{#1}}\fi
\expandafter\ifx\csname urlprefix\endcsname\relax\def\urlprefix{URL }\fi
\expandafter\ifx\csname href\endcsname\relax
  \def\href#1#2{#2} \def\path#1{#1}\fi

\bibitem{adamatzky2021fungal}
A.~Adamatzky, A.~Gandia, A.~Chiolerio, Fungal sensing skin, Fungal biology and
  biotechnology 8~(1) (2021) 1--6.

\bibitem{karana2018material}
E.~Karana, D.~Blauwhoff, E.-J. Hultink, S.~Camere, When the material grows: A
  case study on designing (with) mycelium-based materials, International
  Journal of Design 12~(2) (2018).

\bibitem{jones2020engineered}
M.~Jones, A.~Mautner, S.~Luenco, A.~Bismarck, S.~John, Engineered mycelium
  composite construction materials from fungal biorefineries: A critical
  review, Materials \& Design 187 (2020) 108397.

\bibitem{cerimi2019fungi}
K.~Cerimi, K.~C. Akkaya, C.~Pohl, B.~Schmidt, P.~Neubauer, Fungi as source for
  new bio-based materials: a patent review, Fungal biology and biotechnology
  6~(1) (2019) 1--10.

\bibitem{pelletier2013evaluation}
M.~Pelletier, G.~Holt, J.~Wanjura, E.~Bayer, G.~McIntyre, An evaluation study
  of mycelium based acoustic absorbers grown on agricultural by-product
  substrates, Industrial Crops and Products 51 (2013) 480--485.

\bibitem{elsacker2020comprehensive}
E.~Elsacker, S.~Vandelook, A.~Van~Wylick, J.~Ruytinx, L.~De~Laet, E.~Peeters, A
  comprehensive framework for the production of mycelium-based lignocellulosic
  composites, Science of The Total Environment 725 (2020) 138431.

\bibitem{dias2021investigation}
P.~P. Dias, L.~B. Jayasinghe, D.~Waldmann, Investigation of mycelium-miscanthus
  composites as building insulation material, Results in Materials 10 (2021)
  100189.

\bibitem{wang2016experimental}
F.~WANG, H.-q. LI, S.-s. KANG, Y.-f. BAI, G.-z. CHENG, G.-q. ZHANG, The
  experimental study of mycelium/expanded perlite thermal insulation composite
  material for buildings, Science Technology and Engineering 2016 (2016) 20.

\bibitem{cardenas2020thermal}
J.~P. C{\'a}rdenas-R, Thermal insulation biomaterial based on hydrangea
  macrophylla, in: Bio-Based Materials and Biotechnologies for Eco-Efficient
  Construction, Elsevier, 2020, pp. 187--201.

\bibitem{holt2012fungal}
G.~Holt, G.~Mcintyre, D.~Flagg, E.~Bayer, J.~Wanjura, M.~Pelletier, Fungal
  mycelium and cotton plant materials in the manufacture of biodegradable
  molded packaging material: Evaluation study of select blends of cotton
  byproducts, Journal of Biobased Materials and Bioenergy 6~(4) (2012)
  431--439.

\bibitem{mojumdar2021mushroom}
A.~Mojumdar, H.~T. Behera, L.~Ray, Mushroom mycelia-based material: An
  environmental friendly alternative to synthetic packaging, Microbial Polymers
  (2021) 131--141.

\bibitem{adamatzky2019fungal}
A.~Adamatzky, P.~Ayres, G.~Belotti, H.~W{\"o}sten, Fungal architecture position
  paper., International Journal of Unconventional Computing 14 (2019).

\bibitem{adamatzky2021reactive}
A.~Adamatzky, A.~Nikolaidou, A.~Gandia, A.~Chiolerio, M.~M. Dehshibi, Reactive
  fungal wearable, Biosystems 199 (2021) 104304.

\bibitem{silverman2020development}
J.~Silverman, H.~Cao, K.~Cobb, Development of mushroom mycelium composites for
  footwear products, Clothing and Textiles Research Journal 38~(2) (2020)
  119--133.

\bibitem{appels2020use}
F.~V.~W. Appels, The use of fungal mycelium for the production of bio-based
  materials, Ph.D. thesis, Universiteit Utrecht (2020).

\bibitem{jones2020leather}
M.~Jones, A.~Gandia, S.~John, A.~Bismarck, Leather-like material biofabrication
  using fungi, Nature Sustainability (2020) 1--8.

\bibitem{leo}
M.~Cucchi, H.~Kleemann, H.~Tseng, G.~Ciccone, A.~Lee, D.~Pohl, K.~Leo, Directed
  growth of dendritic polymer networks for organic electrochemical transistors
  and artificial synapses, Adv. Electron. Mat. 7 (2021) 2100586.

\bibitem{filament1}
K.~Janzakova, A.~Kumar, M.~Ghazal, A.~Susloparova, Y.~Coffinier, F.~Alibart,
  S.~Pecqueur, Analog programing of conducting-polymer dendritic
  interconnections and control of their morphology, Nat. Commun. 12 (2021)
  6898.

\bibitem{pismen1}
L.~Pismen, Morphogenesis deconstructed. An integrated view of the generation of
  forms, Springer, 2020.

\bibitem{kasai}
S.~Kasai, S.~Inoue, S.~Okamoto, K.~Sasaki, X.~Yin, R.~Kuroda, M.~Sato,
  R.~Wakamiya, K.~Saito, Detection and control of charge state in single
  molecules toward informatics in molecule networks, Springer, 2017.

\bibitem{beasley2022mem}
A.~E. Beasley, M.-S. Abdelouahab, R.~Lozi, M.-A. Tsompanas, A.~L. Powell,
  A.~Adamatzky, Mem-fractive properties of mushrooms, Bioinspiration \&
  Biomimetics 16~(6) (2022) 066026.

\bibitem{adamatzky2021electrical}
A.~Adamatzky, A.~Gandia, On electrical spiking of ganoderma resinaceum,
  Biophysical Reviews and Letters 16~(04) (2021) 133--141.

\bibitem{beasley2020fungal}
A.~E. Beasley, A.~L. Powell, A.~Adamatzky, Fungal photosensors, arXiv preprint
  arXiv:2003.07825 (2020).

\bibitem{adamatzky2022living}
A.~Adamatzky, A.~Gandia, Living mycelium composites discern weights via
  patterns of electrical activity, Journal of Bioresources and Bioproducts
  7~(1) (2022) 26--32.

\bibitem{dehshibi2021stimulating}
M.~M. Dehshibi, A.~Chiolerio, A.~Nikolaidou, R.~Mayne, A.~Gandia,
  M.~Ashtari-Majlan, A.~Adamatzky, Stimulating fungi pleurotus ostreatus with
  hydrocortisone, ACS Biomaterials Science \& Engineering 7~(8) (2021)
  3718--3726.

\bibitem{roberts2021mining}
N.~Roberts, A.~Adamatzky, Mining logical circuits in fungi, arXiv preprint
  arXiv:2108.05336 (2021).

\bibitem{slayman1976action}
C.~L. Slayman, W.~S. Long, D.~Gradmann, ``action potentials” in neurospora
  crassa, a mycelial fungus, Biochimica et Biophysica Acta (BBA)-Biomembranes
  426~(4) (1976) 732--744.

\bibitem{olsson1995action}
S.~Olsson, B.~Hansson, Action potential-like activity found in fungal mycelia
  is sensitive to stimulation, Naturwissenschaften 82~(1) (1995) 30--31.

\bibitem{adamatzky2018spiking}
A.~Adamatzky, On spiking behaviour of oyster fungi pleurotus djamor, Scientific
  reports 8~(1) (2018) 1--7.

\bibitem{cocatre1992identification}
J.~Cocatre-Zilgien, F.~Delcomyn, Identification of bursts in spike trains,
  Journal of neuroscience methods 41~(1) (1992) 19--30.

\bibitem{legendy1985bursts}
C.~Legendy, M.~Salcman, Bursts and recurrences of bursts in the spike trains of
  spontaneously active striate cortex neurons, Journal of neurophysiology
  53~(4) (1985) 926--939.

\bibitem{dehshibi2021electrical}
M.~M. Dehshibi, A.~Adamatzky, Electrical activity of fungi: Spikes detection
  and complexity analysis, Biosystems 203 (2021) 104373.

\bibitem{adamatzky2022language}
A.~Adamatzky, Language of fungi derived from their electrical spiking activity,
  Royal Society Open Science 9~(4) (2022) 211926.

\bibitem{dactron2003understanding}
L.~Dactron, Understanding fft windows, Application Note. lds group (2003).

\bibitem{przyczyna2022}
D.~Przyczyna, G.~Hess, K.~Szaciłowski, {KNOWM} memristors in a bridge synapse
  delay-based reservoir computing system for detection of epileptic seizures,
  Int. J. Parallel Emergent Distrib. Syst. 37~(5) (2022) 512--527.

\bibitem{mendel}
J.~M. Mendel, Fuzzy logic systems for engineering: a tutorial, Proceedings of
  the IEEE 83~(3) (1995) 345--377.

\end{thebibliography}

\end{document}